# On-device Internet of Sounds Sonification with Wavetable Synthesis Techniques for Soil Moisture Monitoring in Water Scarcity Contexts

Stephen Roddy
*Radical Humanities Laboratory &
Department of Digital Humanities,
University College Cork*
Cork, Ireland
sroddy@ucc.ie

*Abstract*— Sonification, the mapping of data to sound to communicate information about the original data source, is becoming a viable strategy for the sonic representation and communication of information derived from the complex flows of data exchanged across Internet of Sounds (IoS) networks. This paper presents an IoS sonification implementation for monitoring soil moisture levels within the broader context of the globally increasing water scarcity. While previous work has focused on sonifications operating on the applications and services level of the IoS network infrastructure, this paper explores device-level sonification using wavetable synthesis techniques to map sensor data to acoustic parameters. An approach to on-device wavetable sonification is formalized, and a prototype implementation is presented and explored before the approach is contextualised with regard to the soil moisture monitoring tasks.

## I. Introduction

Turchet et al. [1] laid out a future roadmap for research in the Internet of Sounds, establishing some basic schemas for IoS networks and calling for, amongst other things, a deeper integration of sonification techniques in IoS contexts. Roddy [2] put forward a sonification-enabled IoS network schema that integrated sonification techniques into the application layer and device layer of Turchet et al.'s original IoS network schema for applications across a diverse range of scenarios. The accompanying implementation focused on an IoS sonification solution for the monitoring and analysis of Smart City data at the application layer of the IoS network schema. The work described here extends sonification to the IoS network schema's edge, exploring how sonification might be integrated at the device level of a network for environmental monitoring tasks aimed at addressing the growing problem of global water scarcity.

The advantage of on-device sonification running at the edge of an IoS network is that those stakeholders monitoring or carrying out tasks with the devices can access the data on the device via sonification. Thus, they can address problems in or related to the measured phenomenon as indicated in the sonification in which the sensor data is encoded. This differs from sonification at the application layer of a network, which is often geographically and temporally isolated from the site and timeframe in which the data was collected. Tasks at this level tend to focus instead on monitoring and analysis across subsets of interests determined from the aggregate of devices that compose the network edge. While individual sensor values are accessible at the applications layer, it is important that devices also make their data available at the edge for contexts where stakeholders need to monitor and react to sensor values, logged or live, in real-time.

## II. Global Water Scarcity & IoS Sonification

Water scarcity is an increasingly common problem worldwide as a global water crisis, driven by a range of factors from climate change to sub-standard infrastructural and distribution systems, continues to unfold [3]. 4 billion people worldwide experience severe water scarcity for at least one month out of every year [4] and there are 3.2 billion people living in agricultural areas which experience high degrees of water scarcity [5], with water scarcity expected to displace 700 million people by 2030 [6]. A key contributing factor to water scarcity is the global agricultural industry, which is responsible for 69% of global water withdrawal for irrigation, livestock, and aquaculture activities [7].

A number of IoT-based systems for the monitoring and management of water resources against a backdrop of increasing water scarcity have been proposed in the IoT literature [8], [9], [10], [11], [12]. These applications focus on the management of water resources and the reduction of water waste in an IoT context. The IoS literature, however, has mainly considered water usage from the angle of sustainability in the production and rollout of IoS network resources [13], [14]. The development of IoS solutions that address the water scarcity crisis thus represents a novel development for the field. Pauletto [15] provides a thorough overview and analysis of sonification projects that involve a sustainability component either in their function, where they are intended to aid in sustainability efforts, or in their design, where they are produced with sustainability as a key factor in their production.

There have been a number of sonification systems focused on monitoring water usage in order to track resource consumption [16], to understand the impact of drought [17], to represent agriculturally driven wetlands transformation [18], and finally to better understand the water cycle [19] and its relationship to water scarcity [20]. MIDI Sprout [21] and Plant Wave [22] are hardware devices designed to convert the biological signals produced by plants into music. A number of artists are working with similar devices to turn plant biosignals into music [23]. While Soil Choir [24] was an art installation by Jiří Suchánek that sonified soil moisture in a purely artistic context, the artscience project WeatherChimes [19] is presented as an IoT weather station and sonification system. Although both implementations track and sonify soil moisture data, WeatherChimes is more focused on monitoring environmental variables and analysing trends in the recorded data. The implementation described here builds upon this previous work to present a novel method for sonifying soil moisture data on the device layer of an IoS network for monitoring and analysis tasks. Tracking soil moisture levels in agricultural and horticultural contexts is critical to the good stewardship and management of water resources as it supports the optimisation of irrigation processes and systems for more efficient water usage, thus supporting water conservation. The prototype presented here can help to avoid overwatering by ensuring that plants receive just enough water during watering.

III. PROTOTYPE DEVELOPMENT

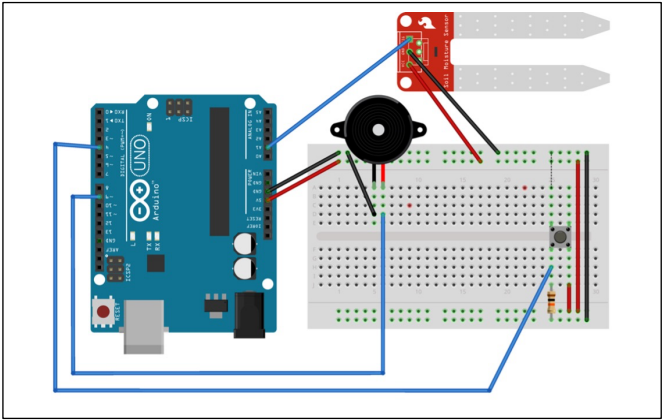

Fig. 1. Initial Prototype Schematic

As per Fig 1, the first prototype was designed using the Arduino Uno R3, a push-button to activate the sonification, a piezo buzzer to play back the sonification and a simple MH-series soil moisture hygrometer which captures moisture data as a drop in restively across the sensor probes in the style of a classic voltage divider, wherein increased soil moisture coinciding with lower resistivity and therefore a higher voltage read. It sonified data live in real-time using FM synthesis techniques. FM synthesis was chosen as previous work found it to be effective for data sonification in IoS contexts where it satisfied the requirements of Poret et. al's formalisation of sonification, (1). which defines sonification $s$ as a function of time $t$, data $X$, acoustic parameters $\theta$, and interaction $I$ in which data $X$ are mapped to meaningful perceptual units $\phi$ further comprised of vectorised sonic dimensions $\vec{p}_i$ ($i \in [1, M]$).

$$s(t, X, \theta, T) = \sum_{i=1}^{M} \phi_i \left( t, \vec{p}_i(t, X(t), \theta(t), T(t)) \right) \quad (1)$$

Rolf Oldeman's [25] FM-synthesis implementation for Uno boards in C/C++ was adapted to sonify the incoming temperature data. While the original plan was to exploit the FM sidebands to map data, it was decided instead to use a simple pitch mapping strategy where increases in sensor value are mapped to increases in pitch, as the audio fidelity was not good enough to support the discrimination of data values from sideband partials. Incoming voltage levels in a range up to 5V are quantised with a 10-bit ADC for 1024 possible values, which are remapped in a range of 50-1000 with these increments added to the fundamental frequency of the FM synthesis routine. The sound synthesis techniques employed push the ATMEGA328P to its absolute limit and manage to produce a single channel of mono audio at a sampling rate of 31.25kHz and 9-bit resolution. This low-fidelity audio is then played back over a piezo buzzer, which is far from an optimal solution given the severe limitations of these components for audio diffusion. Switching the buzzer out for an FR7, a small (6.5cm) 5W 4 Ohm speaker, addresses the limitation of the piezo buzzer but not the limitations of the chip. Furthermore, it is not realistic to expect the speaker to be a reliable method of presenting the data in a noisy environment. Depending on conditions, it may be very difficult to hear the audio signals. The FM synthesis approach, while proving useful for representing multiple streams of highly variable data, is not needed for an application of this nature, where there is only one single data variable to be sonified at any given time. However, the capacity of FM synthesis to represent changes in data as changes in the perceived timbre of the sound was critical to the operation of the previous IoS sonification project [2] and, as such, a technique that allows for similar timbral control. The soil moisture sensor was effective and reliable, though given its small form factor and limited probe size of 6cm, it is best suited to monitoring soil moisture levels for shallow-rooting plants. This point is explored in greater detail shortly.

*B. Second Prototype*

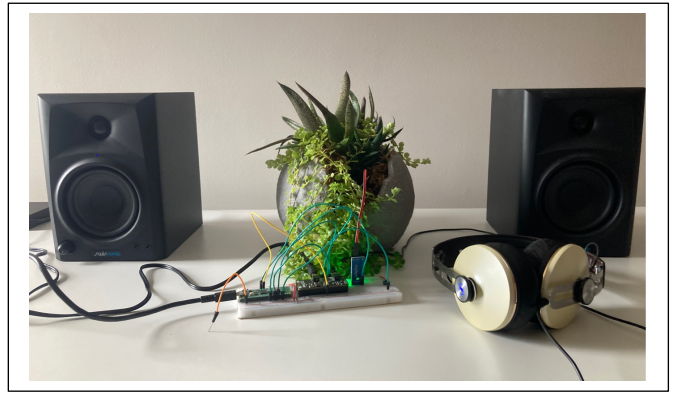

Fig. 2. Second Prototype

A second prototype, Fig. 2 above, was developed to address the limitations encountered in the design of the initial prototype while introducing a facility for sonifying logged data on the device's flash memory. The design was specified to produce high-quality audio, a more reliable method of delivering that audio to a user, and a more relevant sound synthesis technique for representing and communicating the data in question. As per Fig. 3, this prototype is built on the Raspberry Pi Pico 1 with audio amplification and output handled with the Pimoroni Pico Audio Pack shield [26]. This shield features a PCM5100A DAC capable of up to 32-bit stereo audio at a sample rate of 384KHz. In practice, the sample rate and bit depth are constrained by the hardware limitations of the Pico, which produces 16-bit stereo audio at 48kHz. This is nonetheless a dramatic increase in audio fidelity over the ATMEGA328P-based prototype, which, as per above, could handle a single mono channel at a 31.25kHz sample rate at a 9-bit resolution. Furthermore, the board includes a PAM8908JER stereo headphone amplifier, which allows the user to listen to the sonification over headphones, eliminating the need for speakers. The PCM5100A is capable of up to 32-bit stereo audio at a sample rate of 384KHz, the Pico is synthesising 16-bit stereo audio at 48kHz per second.

Audio data is communicated from the Pico to the audio shield over wire via the I2S protocol. I2S uses pulse-code modulation (PCM) to exchange 2-channel stereo audio over a serial interface between integrated circuit (IC) components using a clock, data, and channel select lines. This stereo capability is exploited in the sonification mapping strategy discussed shortly. The sensor remains the same as in the first prototype, an MH-series soil moisture hygrometer. Where the initial prototype was written in C/C++ the code for this implementation is written in Adafruit's branch of Micropython for embedded systems: CircuitPython 9.2.7 [27]. Being an interpreted language, CircuitPython runs far slower than compiled C/C++ might. However, this prototype makes extensive use of CircuitPython's ulab [28] and synthio [29] libraries for sound synthesis, both of which are written in C/C++ and therefore run much faster and more efficiently than we might expect from a purely interpreted language.

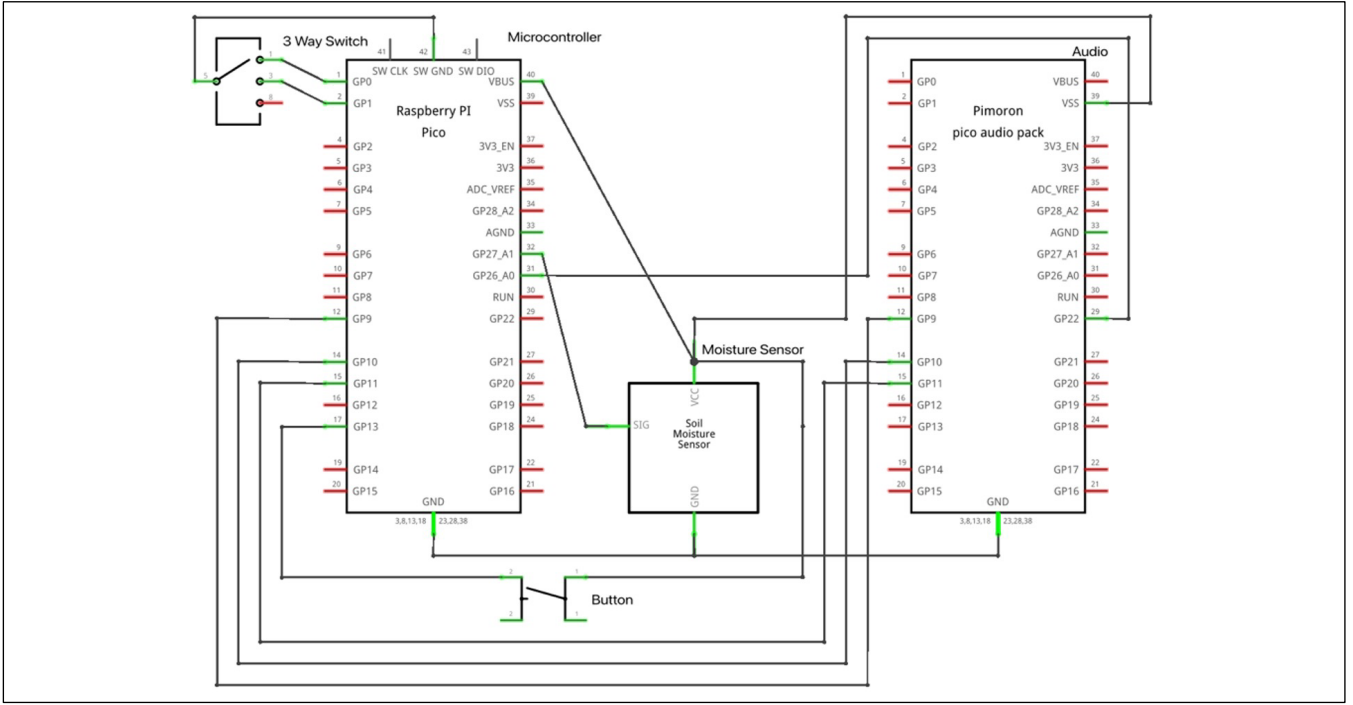

Fig. 3. Second Prototype Schematic

The prototype has three modes of operation. The first is a datalogging mode in which all sensor readings are logged to a text file on flash memory at a rate of 1Hz. This is the standard mode in which we would expect a traditional IoT device to operate. While the pico has about 2MB of flash memory on board, this is easily extendable through the addition of an SD/microSD card module. Live sonification mode takes incoming values polled from the sensor and maps them directly to audio parameters in real-time. This mode is intended for real-time monitoring of soil moisture levels. Finally, the historical sonification mode takes the entire sequence of logged sensor values recorded on the device's flash memory and maps them to sound synthesis parameters, playing them all back in sequence, starting with the earliest recorded data point. Switching between read and write modes is handled by a boot script and thus requires a reboot, which enables CircuitPython to write sensor values to the filesystem in datalogging mode, and to read them from the filesystem in either of the two sonification modes.

### C. Wavetable Synthesis

Wavetable synthesis was chosen as the method for sonifying the data over FM synthesis. It was chosen because it allows data to be mapped to control the complex timbral characteristics of a sound as well as the usual acoustic parameters of pitch, duration, and amplitude popular in parameter mapping sonification (PMSON) [30]. Digital wavetable synthesis was pioneered independently by Wolfgang Palm [31] and David McNabb in the late 1970s, building on Max Mathews [32] implementation for analog computing in his MUSIC II software in 1958. The digital approach to wavetable synthesis generally involves the storage of single-cycle waveforms of differing waveshapes in tables, between which interpolation can be performed to produce novel audio signals [33]. The details of this technique as they relate to sonification are explored in greater detail below. Wavetable synthesis has proven a useful technique for auditory display contexts, being specified and recommended as such by Cook in his exploration of sound synthesis techniques for sonification [30]. More recently, Bergren et al. [34] have applied wavetable synthesis techniques to the analysis of graphene optoelectronics data. It similarly applied Joo's [35] and Horsak et al. [36] in the contexts of colour and gait data, respectively, while the associated technique of wave space synthesis is applied to sonification by Hermann [37] and Kacem et al. [38].

### D. Formalizing a Wavetable Synthesis Sonification Strategy

$$y(t) = A\sin(2\pi f t + \phi) \qquad (2)$$

The audio waveforms are derived and generated from the formulae for their respective waveshapes and stored in ulab arrays that operate as wavetables. Each cycle is 512 samples in length. This implementation uses a sine wave cycle as per (2) and a sawtooth waveform cycle (falling) as per (3).

$$x(t) = -\frac{2a}{\pi}\sum_{n=1}^{N}(-1)^n\frac{\sin(2\pi n f t + \phi)}{n} \qquad (3)$$

Wavetable synthesis is implemented as laid out in (4) by buffering the results of a linear interpolation between commonly indexed audio samples. S is the interpolated audio sample between samples *S1* and *S2* from the first and second wavetables, respectively. The parameter *k* is the index correlating to the interpolated sample S that sits between i1 and *i2*, the indices for *S1*, and *S2*, respectively, an approach described by Masie [34] as 2-point interpolation.

$$S = S1 + (k - i1)(S2 - S1) \qquad (4)$$

We can use the interpolation parameter $k$ ($k \in R$, $0 \geq d \leq 1$) to adjust the ratio between interpolated signals (*A:B*) as illustrated in (4). In this implementation, the sensor data, *d* (*d* $\in R$, $0 \geq d \leq 1$) in (5) and Fig. 5 is substituted for this parameter (*k*), allowing the moisture levels in the soil to directly control the interpolation and thus inform the timbre of the sound.

$$X(t) = a_i + (1 - dt) + b_i(dt) \qquad (5)$$

This is represented by the *terms 1-dt* and *dt* in (5) and Fig. 5 where a and b represent the current output samples from wavetables A and B respectively. Each wavetable consists of three oscillators, denoted by *i* (*i* $\in W$, *i* $\leq 2$), with unique pitch frequencies, $f_{i=0}$, $f_{i=1}$ and $f_{i=2}$ which bare harmonic relation to one another as root, third, and fifth notes respectively.

$$F(t) = r + f_j t - rd \qquad (6)$$

The frequencies of these pitches are determined by the data as illustrated in (6), in which *d* (*d* $\in R$, $0 \geq d \leq 1$) represents the data, and *r* (*r* $\in W$, *r* $\leq m$) is the frequency range to which the data is mapped. The frequency of the highest partial is represented by *m*. The pitch mapping strategy here is inverted, as per the *r-rd* term, meaning that as the moisture value increases, the pitch drops and vice versa. The exact ranges involved in the pitch mapping are outlined in Table 1.

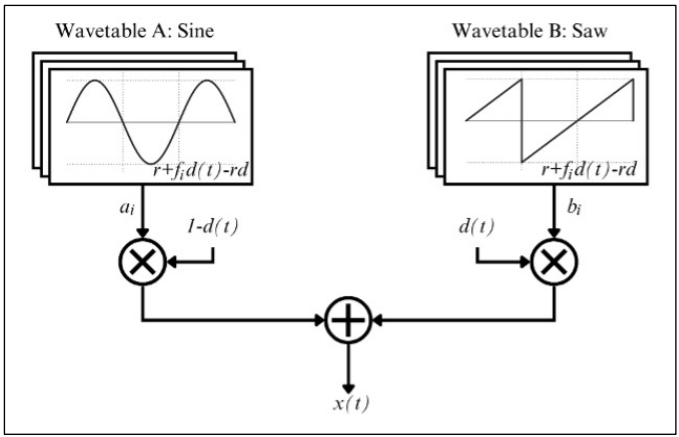

Fig. 4. Wavetable Synthesis Sonification Strategy

## IV. IMPLEMENTING THE SONIFICATION

### A. Mapping Parameters

Both wavetable synthesisers have their outputs passed to filters and amplitude envelopes intended to shape and refine the sound. Both wavetable routines produce a single three-note chordal structure. Note information is coded in the MIDI protocol for ease, and then converted to frequency data for sonification. As such, the pitch mapping strategy is somewhat nonlinear from a frequency perspective, though it remains linear from a musical point of view. Recalling the previous formulae, data is mapped to control two distinct components of the sounds generated by the wavetables. The first is a pitch offset subtracted from note values an octave above the root, third, and fifth of each chord, thus producing an octave range as generalised in (6) previously and for which the exact ranges are presented in table 1 below. This is an inverse polarity mapping in which the pitch is decremented as soil moisture values increase.

TABLE I. PITCH MAPPING PARAMATERS

| Note | MIDI | Pitch | Frequency |
|---|---|---|---|
| **Low Parameters** | | | |
| Root | 55 | G3 | 195Hz |
| Third | 59 | B3 | 246.94Hz |
| Fifth | 62 | D4 | 293.66Hz |
| **High Parameters** | | | |
| Root | 67 | G4 | 392Hz |
| Third | 71 | B4 | 493.88Hz |
| Fifth | 74 | D5 | 587.33Hz |

The second is the k-parameter that interpolates between the two waveshapes and thus controls the shape of the resulting wave as described in (4) above. When producing a sonification by performing 2-point interpolative wavetable synthesis between sine and sawtooth waves we are controlling, as a single group, the amplitudes of all harmonics N above the fundamental. In the case of the current mapping strategy, we are boosting these harmonics as the soil moisture values increase, and attenuating them as the data decreases. The perceived effect here is different to the successive addition of harmonic partials in sequence which may mistakenly suggest to the listener that each additional partial is representative of a specific change in the data. Rather this approach builds upon what worked in prototype 1 and in previous work [2] by redundantly mapping the data to control the spectral shape and complexity of the sound. The rationale behind the mapping strategy is motivated by research in timbral perception that suggests sine waves are perceived as smooth and dark while sawtooth waves are heard as bright and rough [39], concepts which naturally map onto representations of moisture and dryness respectively.

In the live sonification mode, the chord plays for as long as the button is pressed, while in the logged data sonification mode, each successive data point is played as a single impulse at a rate of one per second. The signal chain involved is illustrated in Fig. 5. Both modes employ a gentle amplitude envelope that shapes the sound into a slowly evolving pulse shape with a 200ms attack and an 800ms decay. Each note is further processed with a resonant low-pass filter with a cut-off frequency of 4000kHz and a Q factor of .5. These components create a warm, somewhat 'blurry' tone that dies out slowly as the note is released. Each root note is panned centrally in the stereo stage while the third and fifth notes are panned 45° to the left and right, respectively. Spatialising the chordal shape across stereo space in this manner helps to reduce masking and ensure that the notes are clearly conveyed to the listener. The same filter, envelope, and panning parameters are used for both the live sonification and the sonification of previously logged data.

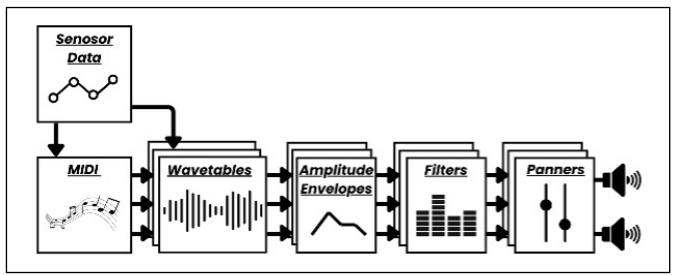

Fig. 5. Signal Chain for Wavetable Sonification

When the button is held down in live sonification mode, the initial chordal configuration will correspond to the soil moisture level. If the moisture changes while the button is held, those chords will remain the same but the timbre of the sound, given by the degree of interpolation between the saw and sine waveforms, will change. This allows a user to track soil moisture during watering as a function of the change in timbre. Because the initial chord is held as long as the button is held, the user is provided with a measure of the initial soil moisture against which they can compare the representation of the current soil moisture level as encoded on the timbre. This is useful for monitoring in real-time as the plants are being watered. Examples of the system in operation, along with the code and further details are available at the link here [40].

*B. Addressing Aliasing*

Aliasing is a problem for wavetable synthesis where high-frequency content produced by the interpolation of two samples can surpass the Nyquist limit and thus cause distortion and spectral wrap-around. This is especially true when working with sawtooth waves, which are comprised of a large number of harmonic partials above the fundamental. One solution to this problem is oversampling, which simply involves increasing the sampling rate to account for all of the high-frequency partials in the sound. However, the very high sample rates required to avoid aliasing in this context would be far in excess of the capabilities of the device hardware used for this implementation.

The better approach here is to break up the frequency spectrum into bandlimited subtables where, for example, each octave (or doubling of frequency in the spectrum) uses a uniquely bandlimited wavetable oscillator that produces tones with only half the harmonics of the previous subtable. This way, the number of partials decreases as the fundamental frequency increases, and thus the Nyquist limit is never crossed. This approach has generally been pursued by instrument designers who need to ensure that their synthesisers are capable of playing across a full musical octave range, and is not necessarily relevant for the implementation of sonification presented here, where we are working within one single octave range [33].

It is possible to approach wavetable synthesis with sawtooth waves in Synthio by creating a simple rising or falling line in an ulab array with 512 index positions, one per sample, where each sample is encoded as a signed 16-bit integer. The synthio engine will then loop these waveshapes to synthesise sound. The problem with this approach is that it does not allow the developer to specify the number of harmonic partials in the saw wave, and as a result, aliasing can quite easily occur, producing quite prominent wraparound artifacts as the Nyquist limit is exceeded. This project adopts an additive approach to the production of the sawtooth cycle, creating the waveshape by successively adding sine wave partials N at harmonic intervals over a fundamental. This allows for control of the harmonic content of the saw waves. The fundamental frequencies of our pitches stay within a range of 196Hz to 587.33Hz. Our highest partial then, when $N = 17$ will be 9984.61Hz, thus requiring a sample rate of 19,969.22Hz. This is below our sample rate of 48,000Hz, thereby getting the highest number of possible frequency components while avoiding any aliasing artefacts. This is implemented by ensuring that each component of the saw wave is encoded within a range of the bit depth divided by N.

V. EVALUATION

A preliminary user evaluation (n=9) was conducted to gather some data on the efficacy of the system. Six participants identified as female and 3 as male. 3 were aged between 18-24, 1 between 25-34, 2 in the rage of 35-44 and 45-54 respectively and one as in the 55-64 age range. All but one participants reported normal hearing with how reporting a suspected (but undiagnosed) deficiency in one or both ears, with 7 participants reporting varying degrees of musical education. Users took part in a short demonstration session to familiarise them with the prototype and its operation. They were presented with the prototype and two potted plants, a Ficus Pumila 'Green sunny' that had been recently watered, and thus had higher soil moisture levels, and a Crassula Marginalis which had much dryer soil. Users were shown how to safely and effectively insert the sensor probe into the soil for each plant. They were then asked to use the prototype to take readings from both plants, listening to the sonified data. The device was then booted in historical sonification mode, and users listened to a sonification of 108 data values representing slowly decreasing moisture levels. Users were then presented with an evaluation survey consisting of two sections. The first involved 6 questions intended to determine whether users could identify the dry plant from the wet plant and gauge the level of agreement between users on the degree of moisture present in the two plant soils. This was followed by a BUZZ scale test [41]. The 11 statements of the BUZZ scale are scored from 1-7 and when taken together, they provide information on the effectiveness and efficacy of an auditory display.

*A. Results & Interpretation*

6 users correctly identified the plant with the moist, recently watered soil in Q1 while in Q2 users rated the moisture levels of soil on a Likert scale from 1=Very Dry to 5=Very Wet ($\mu = 3.44$, SD = 1.5). These results, while not very strong, suggest a fair deal of agreement between listeners on the moisture levels present. In Q3, 6 users also identified the plant with the dry soil with users rating the moisture levels on a Likert scale from 1=Very Dry to 5=Very Wet ($\mu = 2.44$, SD = 1.74) in Q4. Again these results would seem to imply some agreement between listeners on the soil moisture levels. In the historical sonification (Q5), 6 users correctly identified that the moisture levels were falling and in Q6 they scored the rate of change in these moisture levels on a scale from 1=Very Slowly to 5=Very Quickly ($\mu = 2.77$, SD = 1.09) suggesting a good deal of agreement on the rate of change. The BUZZ scale test results ($\mu = 52.66$, SD = 3.04) provide an overall mean score of 68.39% suggesting a good degree of effectiveness and efficacy, but with substantial room to improve, and the SD further suggests a moderate level of disagreement between users.

## VI. Discussion and Contextualisation

While the evaluation results presented here are preliminary in nature, and limited by the small number of participants involved (n=9), they do nonetheless suggest a certain baseline of efficacy that can be built upon and improved in future work. Such future work will need to focus on ensuring that ambiguity in the sonification signal is minimised so that the data is clearly identifiable for users. This may be achievable through the design of more communicatively effective data to sound mapping strategies.

As global water scarcity becomes an increasingly urgent problem, the rapidly developing IoS ecosystem is positioned to support innovations and solutions that might help to address some of the causes underlying this challenge, as well as the effects produced by it. The device prototype presented and explored here is one such means of addressing the problem in contexts where the overwatering of plants leads to water wastage.

In practice, each species of plant is different with unique root distribution patterns that determine how their root systems are organised and spread throughout the soil in terms of depth, complexity, and span. The device explored here is best suited to monitoring soil moisture conditions for shallow-rooting plant life, which takes root closer to the surface of the soil, thus allowing for the probe prongs of 6cm long to take accurate readings. Water may be unevenly distributed throughout a given body of soil. This problem can be exacerbated by issues like poor soil structure, where bad soil condition and/or a lack of organic materials interferes with the even distribution of water. It is imperative that the sensor takes readings from the same area of the soil from which the root system in question is extracting water. As such, the form factor of the sensor limits its maximum penetration depth, which in turn translates to hard limits in terms of application.

There is, however, quite a large range of vegetation with shallow root systems. For example, in an agricultural context, nuts, berries, fruits, and vegetables with shallow roots include peanuts, strawberries, blackcurrants, scallions, chives, butternut squash, oregano, thyme, and certain kinds of tomato and radish [42]. While from a horticultural point of view, popular shallow-rooted plants include the Ficus Pumila and the Crassula Marginalis used in the evaluation along with azaleas, violas, rhododendrons, petunias, hydrangeas, gardenias, yarrow, aurinia saxatilis, eremurus, and aloe vera [43].

Soil texture becomes a key factor when interpreting the sonified signals produced by the device. Wilting points, the minimum amount of water under which plant life begins to die, available water, the volume of water plants are capable of accessing, and field capacity, the amount of water held in soil after the excess has drained, are highly variable across different categories of soil textures. For example, clay type soils have (% water by volume) ratings of 27.2, 39.6 and 12. for wilting point, field capacity and available water respectively while at the other end of the spectrum soils with a sand type texture have ratings of 3.3, 9.1, and 5.8% across these same measures [44]. As such, the sound produced by the sonification means something very different for plants in clay soils when compared to those that might grow in soils with a sandy texture. This needs to be taken into account by the listener when interpreting the sonification, as the audible result is always relative to the characteristics of the soil in question and is further complicated by the water requirements of the specific vegetation growing therein, which tend to be highly variable across species. As per the mapping strategy previously elucidated, dry soils result in a high G major chord in the 4th octave with a buzzy timbre, while a 3rd octave G major with smoother timbre can be expected from soil with a high moisture content. The midpoint of this scale is a 3rd octave C major chord. With these harmonic and timbral ranges in mind, a listener is in a position to make judgements about the moisture levels in a given body of soil.

However, there is another factor at play here, in that different soil compositions with similar moisture levels will produce results in a drastically variable range on the sensor. As such, the sensor needs to be carefully calibrated for each body of soil with which it is used. This is implementable in CircuitPython when processing the sensor values by observing the maximum and minimum values for a given soil composition and then remapping incoming values to exploit the full range of the sonification mapping strategy as described previously. Calibrating the sensors in this way ensures that the sonified data meaningfully represents the sensor values for a given soil composition and is thus critical to helping manage water resources.

As referenced previously, aliasing, though obviously not specific to wavetable synthesis, provides a particular challenge for the technique. In the 1980s, PPG's Wave series of synthesisers made use of heavy oversampling, which resulted in harmonic imaging, whereby the spectrum was duplicated at harmonics of the sample rate. Being harmonically related and not unpleasant to listen to, these signal components were left in and became a key aesthetic component of PPG's approach to wavetable synthesis [45]. More recent software-based recreations of PPG's wavetable synthesisers have focused on accurately recreating those imaging patterns to better match the aesthetics of the original systems [46].

When carrying out interpolation between two audio samples, as is the case in software-based wavetable synthesis, aliasing, as opposed to harmonic imaging, becomes the problem. However, the argument for leaving these inharmonically related partials in the signal does not hold for the current sonification context, which is typified by information-bearing signals wherein each component can be mapped to a state or value measured from the original data source. In this context, aliasing serves to obscure the information encoded in the signal in much the same way the addition of additional tones over a speech signal would make it more difficult to parse the meaning of the spoken words. As such, it is critical that aliasing is properly addressed and accounted for when designing sonification solutions, and is tackled here by keeping the number of partials in the saw wavetable below the Nyquist limit.

## References

[1] L. Turchet et al., "The Internet of Sounds: Convergent Trends, Insights, and Future Directions," *IEEE Internet of Things Journal*, vol. 10, no. 13, pp. 11264–11292, July 2023, doi: 10.1109/JIOT.2023.3253602.

[2] S. Roddy, "Designing an Internet of Sounds Sonification System with FM Synthesis Techniques," *IEEE Communications Magazine*, vol. 62, no. 12, pp. 42–47, 2024, Accessed: Mar. 20, 2025.

[3] "Water Scarcity," UN-Water. Accessed: May 15, 2025. [Online]. Available: https://www.unwater.org/water-facts/water-scarcity

[4] M. M. Mekonnen and A. Y. Hoekstra, "Four billion people facing severe water scarcity," *Sci. Adv.*, vol. 2, no. 2, p. e1500323, Feb. 2016, doi: 10.1126/sciadv.1500323.

[5] *The State of Food and Agriculture 2020*. FAO, 2020. doi: 10.4060/cb1447en.

[6] "High-Level Panel on 'Harnessing Global Development Agendas on the Road to 2023' during the World Water Week | General Assembly of the United Nations." Accessed: May 15, 2025. [Online]. Available: https://www.un.org/pga/76/2022/08/30/high-level-panel-on-harnessing-global-development-agendas-on-the-road-to-2023-during-the-world-water-week/

[7] Food and Agricultural Organisztion of the United Nations, "AQUASTAT - FAO's Global Information System on Water and Agriculture," fao.org. Accessed: July 06, 2025. [Online]. Available: https://www.fao.org/aquastat/en/overview/methodology/water-use

[8] C. Rajurkar, S. R. S. Prabaharan, and S. Muthulakshmi, "IoT based water management," in *2017 International Conference on Nextgen Electronic Technologies: Silicon to Software (ICNETS2)*, Mar. 2017, pp. 255–259. doi: 10.1109/ICNETS2.2017.8067943.

[9] K. Pachiappan, K. Anitha, R. Pitchai, S. Sangeetha, T. V. V. Satyanarayana, and S. Boopathi, "Intelligent Machines, IoT, and AI in Revolutionizing Agriculture for Water Processing," in *Handbook of Research on AI and ML for Intelligent Machines and Systems*, IGI Global Scientific Publishing, 2024, pp. 374–399. doi: 10.4018/978-1-6684-9999-3.ch015.

[10] S. Mehta and A. Aneja, "Smart Water Management in Agriculture: IoT Solutions for Reducing Water Consumption," in *2024 2nd International Conference on Recent Trends in Microelectronics, Automation, Computing and Communications Systems (ICMACC)*, Dec. 2024, pp. 19–23. doi: 10.1109/ICMACC62921.2024.10894488.

[11] A. Suryavanshi, S. Mehta, S. Chattopadhyay, and M. Aeri, "Navigating Water Scarcity with IoT: A Smart Management System Approach," in *2024 4th International Conference on Intelligent Technologies (CONIT)*, June 2024, pp. 1–5. doi: 10.1109/CONIT61985.2024.10627282.

[12] Upasana et al., "Utilizing Smart Farming Methods to Reduce Water Scarcity," in *Data Driven Mathematical Modeling in Agriculture*, River Publishers, 2024.

[13] L. Gabrielli and L. Turchet, "Towards a Sustainable Internet of Sounds," in *AudioMostly 2022*, St. Pölten Austria: ACM, Sept. 2022, pp. 231–238. doi: 10.1145/3561212.3561246.

[14] L. Gabrielli, E. Principi, and L. Turchet, "Sustainability and the Internet of Sounds: Case Studies," *IEEE Transactions on Technology and Society*, pp. 1–16, 2024, doi: 10.1109/TTS.2024.3513777.

[15] S. Pauletto, "Sonification and sustainability," in *The Routledge Handbook of Sound Design*, 1st ed., London: Focal Press, 2024, pp. 304–317. doi: 10.4324/9781003325567-21.

[16] Y. Seznec and S. Pauletto, "The Singing Shower: A Melody-Sensitive Interface For Physical Interaction and Efficient Energy Consumption," June 2022, doi: 10.5281/ZENODO.6573519.

[17] Y. C. Han, "California Drought Impact v2: Data Visualization and Sonification using Advanced Multimodal Interaction," in *Proceedings of the 2017 CHI Conference Extended Abstracts on Human Factors in Computing Systems*, in CHI EA '17. New York, NY, USA: Association for Computing Machinery, May 2017, pp. 1372–1377. doi: 10.1145/3027063.3052542.

[18] D. G. Angeler, M. Alvarez-Cobelas, and S. Sánchez-Carrillo, "Sonifying social-ecological change: A wetland laments agricultural transformation," *E&S*, vol. 23, no. 2, p. art20, 2018, doi: 10.5751/ES-10055-230220.

[19] W. Woo, W. Richards, J. Selker, and C. Udell, "WeatherChimes: An Open IoT Weather Station and Data Sonification System," *HardwareX*, vol. 13, p. e00402, Mar. 2023, doi: 10.1016/j.ohx.2023.e00402.

[20] L. Rustad, X. Cortada, M. Quinn, and T. Hallett, "The Waterviz for Hubbard Brook: The Confluence of Science, Art, and Music at Long-Term Ecological Research Sites," in *Field to Palette*, CRC Press, 2018.

[21] "MIDI Sprout," MIDI Sprout. Accessed: May 19, 2025. [Online]. Available: https://www.midisprout.com

[22] "PlantWave," PlantWave. Accessed: May 19, 2025. [Online]. Available: https://plantwave.com/en-ie

[23] "Meet the electronic musicians collaborating with nature through bio-sonification," DJ Mag. Accessed: May 19, 2025. [Online]. Available: https://djmag.com/features/meet-electronic-musicians-collaborating-nature-through-bio-sonification

[24] J. Suchánek, "SOIL CHOIR v.1.3 - soil moisture sonification installation," in *Proceedings of the International Conference on New Interfaces for Musical Expression*, Zenodo, June 01, 2020, pp. 617--618. doi: 10.5281/zenodo.4813226.

[25] rgco, "Arduino Synthesizer With FM," Instructables. Accessed: May 25, 2025. [Online]. Available: https://www.instructables.com/Arduino-Synthesizer-With-FM/

[26] "Pico Audio Pack (Line-Out and Headphone Amp)." Accessed: May 25, 2025. [Online]. Available: https://shop.pimoroni.com/products/pico-audio-pack

[27] "CircuitPython." Accessed: May 23, 2025. [Online]. Available: https://circuitpython.org/

[28] "ulab – Manipulate numeric data similar to numpy — Adafruit CircuitPython 1 documentation." Accessed: May 23, 2025. [Online]. Available: https://docs.circuitpython.org/en/latest/shared-bindings/ulab/

[29] "synthio – Support for multi-channel audio synthesis — Adafruit CircuitPython 1 documentation." Accessed: May 23, 2025. [Online]. Available: https://docs.circuitpython.org/en/latest/shared-bindings/synthio/

[30] P. Cook, "Sound Synthesis for Auditory Display," in *The sonification handbook*, T. Hermann, A. Hunt, J. Neuhoff, and Europäische Zusammenarbeit auf dem Gebiet der Wissenschaftlichen und Technischen Forschung, Eds., Berlin: Logos Verlag, 2011, pp. 197–235.

[31] "Wolfgang Palm - PPG Apps." Accessed: Mar. 02, 2025. [Online]. Available: https://palm.seib.info/story/c4.html

[32] R. Boulanger and V. Lazzarini, *The Audio Programming Book*. MIT Press, 2010.

[33] D. C. Massie, "Wavetable Sampling Synthesis," in *Applications of Digital Signal Processing to Audio and Acoustics*, M. Kahrs and K. Brandenburg, Eds., Boston, MA: Springer US, 2002, pp. 311–341. doi: 10.1007/0-306-47042-X_8.

[34] A. J. Bergren, A. Beltaos, and A. Van Dijk, "Sonification methods for enabling augmented data analysis applied to graphene optoelectronics," *Applied Research*, vol. 3, no. 5, p. e202300092, Oct. 2024, doi: 10.1002/appl.202300092.

[35] W. Joo, "Sonifyd: A Graphical Approach for Sound Synthesis and Synesthetic Visual Expression," in *Proceedings of the 25th International Conference on Auditory Display (ICAD 2019)*, Newcastle upon Tyne: Department of Computer and Information Sciences, Northumbria University, June 2019, pp. 96–102. doi: 10.21785/icad2019.045.

[36] B. Horsak et al., "A WIRELESS INSTRUMENTED INSOLE DEVICE FOR REAL-TIME SONIFICATION OF GAIT," 2015.

[37] T. Hermann, "Wave Space Sonification," in *Proceedings of the 24th International Conference on Auditory Display - ICAD 2018*, Houghton, Michigan: The International Community for Auditory Display, June 2018, pp. 49–56. doi: 10.21785/icad2018.026.

[38] A. Kacem, K. Zbiss, P. Watta, and A. Mohammadi, "Wave space sonification of the folding pathways of protein molecules modeled as hyper-redundant robotic mechanisms," *Multimed Tools Appl*, vol. 83, no. 2, pp. 4929–4949, Jan. 2024, doi: 10.1007/s11042-023-15385-y.

[39] Z. Wallmark, *Nothing But Noise: Timbre and Musical Meaning at the Edge*. Oxford University Press, 2022.

[40] "On-Device Intrnet of Sounds Sonification… - Zenodo." Accessed: Wed 24, 2025. [Online]. Available: https://doi.org/10.5281/zenodo.17193530

[41] B. J. Tomlinson, B. E. Noah, and B. N. Walker, "BUZZ: An Auditory Interface User Experience Scale," in *Extended Abstracts of the 2018 CHI Conference on Human Factors in Computing Systems*, Montreal QC Canada: ACM, Apr. 2018, pp. 1–6. doi: 10.1145/3170427.3188659.

[42] "RHS Grow Fruit & Veg Guide - Dubray Books." Accessed: May 26, 2025. [Online]. Available: https://www.dubraybooks.ie

[43] C. R. Adams, *Principles of Horticulture*, 6th ed. London: Routledge, 2012. doi: 10.4324/9780080969589.

[44] "Estimation of Soil Water Properties", Accessed: May 26, 2025. [Online]. Available: https://doi.org/10.13031/2013.33720

[45] C. Meyer, "The Story of the Prophet VS," Learning Modular. Accessed: May 27, 2025. [Online]. Available: https://learningmodular.com/the-story-of-the-prophet-vs/

[46] SYNTHHEAD, "Waldorf Delivers PPG Wave 3.V," Synthtopia. Accessed: May 27, 2025. [Online]. Available: https://www.synthtopia.com/content/2010/12/08/waldorf-delivers-ppg-wave-3-v/